\documentclass[final,1p,times]{elsarticle}
\usepackage{cases}
\usepackage{amssymb}
\usepackage{amsthm}
\usepackage{lineno}

\newcommand{\ds}{\displaystyle}

\journal{Acta Acustica united with Acustica}

\begin{document}



\begin{frontmatter}

\title{Time-domain numerical modeling of brass instruments including nonlinear wave propagation, viscothermal losses, and lips vibration}

\author[LMA]{Harold Berjamin}
\ead{berjamin@lma.cnrs-mrs.fr}
\author[LMA]{Bruno Lombard}
\ead{lombard@lma.cnrs-mrs.fr}
\author[LMA]{Christophe Vergez}
\ead{vergez@lma.cnrs-mrs.fr}
\author[ECM]{Emmanuel Cottanceau}
\cortext[cor1]{Corresponding author. Tel.: +33 44-84-52-42-53 }
\address[LMA]{Laboratoire de M\'{e}canique et d'Acoustique, UPR 7051 CNRS, 4 impasse Nicolas Tesla, 13453 Marseille Cedex 13, France}


\begin{abstract}
A time-domain numerical modeling of brass instruments is proposed. On one hand, outgoing and incoming waves in the resonator are described by the Menguy-Gilbert model, which incorporates three key issues: nonlinear wave propagation, viscothermal losses, and a variable section. The nonlinear propagation is simulated by a TVD scheme well-suited to non-smooth waves. The fractional derivatives induced by the viscothermal losses are replaced by a set of local-in-time memory variables. A splitting strategy is followed to couple optimally these dedicated methods. On the other hand, the exciter is described by a one-mass model for the lips. The Newmark method is used to integrate the nonlinear ordinary differential equation so-obtained. At each time step, a coupling is performed between the pressure in the tube and the displacement of the lips. Finally, an extensive set of validation tests is successfully completed. In particular, self-sustained oscillations of the lips are simulated by taking into account the nonlinear wave propagation in the tube. Simulations clearly indicate that the nonlinear wave propagation has a major influence on the timbre of the sound, as expected. Moreover, simulations also highlight an influence on playing frequencies, time envelopes and on the playability of the low frequencies in the case of a variable lips tension.\end{abstract}


\end{frontmatter}

\section{Introduction}\label{SecIntro}

Risset and Mathews  \cite{risset1969} were the first to highlight the fact that the spectral enrichment of brass sounds with increasing sound level is crucial to recognize these instruments. They included nonlinear distortion into their additive sound synthesis more than 10 years before acousticians began to focus on this phenomenon, and 25 years before its origin was understood. In 1980, Beauchamp  \cite{beauchamp1980} stressed the fact that a linear model of the air column cannot explain brassy sounds. Since 1996, it is well established that the spectacular spectral enrichment of loud brass sounds is mainly due to the nonlinear wave propagation inside the bore of the instrument  \cite{hirschberg96b, gilbert96, rendon2013}. At extremely high sound levels, shock waves have been observed, but nonlinear distortion even at moderate sound levels can contribute significantly to the timbre of a brass instrument  \cite{campbell2014, norman2010}.

Considering nonlinear propagation is thus fundamental both for sound synthesis by physical modeling  \cite{vergez2000a,msallam2000,helie2008a,Bilbao11} and to improve the understanding of musical instruments design  \cite{myers2012,gilbert2008,chick2012a}. One must account for the nonlinear wave propagation of both outgoing and incoming pressure waves, and not only the outgoing pressure wave as it can be done to simplify the problem  \cite{thompson2001}. Besides the nonlinear wave propagation, other mechanisms need also to be incorporated to describe the physics of brass instruments. First, one must handle a continuous variation of the cross section of the instrument with respect to space. Second and more challenging, one must handle the viscothermal losses resulting from the interaction between the acoustic field and the bore of the instrument.

Gilbert and coauthors proposed an approach to handle these mechanisms in the periodic regime. The harmonic balance method has been applied to cylinders with straight tube \cite{Menguy00} or with varying cross section \cite{These_Menguy}. This approach resulted in the development of a simulation tool for brassiness studies \cite{gilbert2008}. 

The time domain offers a more realistic framework to simulate instruments in playing conditions; in counterpart, it introduces specific difficulties. Non-smooth (and possibly non-unique) waves are obtained, whose numerical approximation is not straightforward  \cite{Godlewski96}. Moreover, the viscothermal losses introduce fractional derivatives in time  \cite{Matignon-These,Matignon08}. These convolution products require to store the past values of the solution, which is highly consuming from a computational point of view. These features (nonlinear propagation, viscothermal losses, varying cross section) have been examined separately in the works of Bilbao  \cite{Bilbao11,Bilbao13}. In particular, a discrete filter was used to simulate the memory effects due to the viscothermal losses in the linear regime. But to our knowledge, the full coupling in the time domain between both the nonlinear propagation, the variable section and the viscothermal losses has never been examined. Proposing a unified and  efficient discretization of all these aspects is the first goal of this paper.

Our second objective is to show how to couple the numerical model of resonator to a classical one-mass model for the lips  \cite{Elliott82}. This coupling allows  the full system to be simulated, including the instrument and the instrumentalist both during steady states and transients. Emphasis is put throughout the paper on the choice of the numerical methods and on their validation.

The paper is organized as follows. Section \ref{SecReso} is devoted to the modeling of the resonator. The acoustic propagation inside the bore of the instrument is described by outgoing and incoming nonlinear simple waves, that interact together only at the extremities of the instrument \cite{gilbert2008}. A so-called diffusive approximation is introduced to provide an efficient discretization of the viscothermal losses. Then the equations are solved numerically by following a splitting strategy, which offers a maximal computational efficiency: the propagative part is solved by a TVD scheme (standard in computational fluid dynamics), and the relaxation part is solved exactly. This approach is validated by a set of test-cases; an application to the determination of the input impedance in the linear case is proposed. Section \ref{SecExc} is devoted to the numerical modeling of the exciter. The coupling between the exciter (air blown through vibrating lips) and the resonator (the instrument) is explored in section \ref{SecExp} through various numerical experiments. They show the possibilities offered by the simulation tool developed, and they  highlight the influence of nonlinear propagation on various aspects of the instrument behavior. Lastly, future lines of research are proposed in section \ref{SecConclu}.


\section{Resonator}\label{SecReso}

\subsection{Physical modeling}\label{SecResoPhys}

\subsubsection{Notations}\label{SecResoPhysNot}

\begin{figure}[h!]
\begin{center}
\includegraphics[scale=0.55]{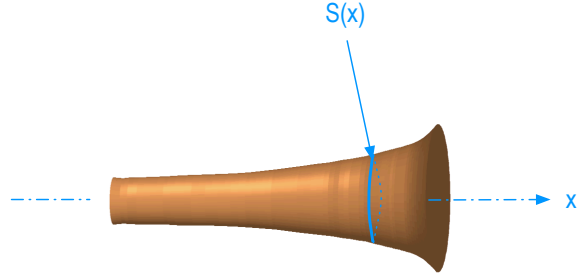} 
\caption{\label{FigGuide1D} One-dimensional acoustic tube of cross section area $S(x)$.}
\end{center}
\end{figure}

A cylinder with radius $R$ depending on the abscissa $x$ is considered. The length of the cylinder is $D$ and its cross section is $S$ (figure~\ref{FigGuide1D}). The physical parameters are the ratio of specific heats at constant pressure and volume $\gamma$; the pressure at equilibrium $p_0$; the density at equilibrium $\rho_0$; the Prandtl number Pr; the kinematic viscosity $\nu$; and the ratio of shear and bulk viscosities $\mu_v/\mu$. One deduces the sound speed $a_0$, the sound diffusivity $\nu_d$ and the coefficient of dissipation in the boundary layer $C$:
\begin{equation}
\begin{array}{l}
\ds
a_0=\sqrt{\frac{\textstyle \gamma\,p_0}{\textstyle \rho_0}}, \hspace{0.2cm}
\nu_d=\nu\left(\frac{\textstyle 4}{\textstyle 3}+\frac{\textstyle \mu_v}{\textstyle \mu}+\frac{\textstyle \gamma-1}{\textstyle \mbox{Pr}}\right),\\
[8pt]
\ds
C=1+\frac{\textstyle \gamma-1}{\textstyle \sqrt{\mbox{Pr}}}.
\end{array}
\label{Omega0}
\end{equation}


\subsubsection{Menguy-Gilbert model}\label{SecResoPhysEqu}

The angular frequency of the disturbance is below the first cut-off angular frequency ($\omega<\omega^*=\frac{1.84\,a_0}{R}$ where $R$ is the maximum radius), so that only the plane mode propagates and the one-dimensional assumption is satisfied \cite{Chaigne08}. Within the framework of weakly nonlinear propagation and assuming that $S$ varies smoothly with $x$, the wave fields are split into simple outgoing waves (denoted $+$) and incoming waves (denoted $-$) that do not interact during their propagation \cite{Hamilton98,These_Menguy}. 

Velocities along the $x$ axis are denoted $u^\pm$. Pressures fluctuations associated with the simple waves are given by
\begin{equation}
p^\pm=\pm \rho_0\,a_0\,u^\pm.
\label{SurP}
\end{equation}
According to the Menguy-Gilbert model, the evolution equations satisfied by the velocities are
\begin{subnumcases}{\label{Chester}}
\displaystyle
\frac{\textstyle \partial u^\pm}{\textstyle \partial t} + \frac{\textstyle \partial}{\textstyle \partial x}\left(\pm au^\pm+b\frac{\textstyle (u^\pm)^2}{\textstyle 2}\right) \pm \frac{\textstyle a}{\textstyle S}\,\frac{\textstyle dS}{\textstyle dx}\,u^\pm \nonumber\\
\displaystyle
\hspace{0.1cm}=\pm c\frac{\textstyle \partial^{-1/2}}{\textstyle \partial t^{-1/2}}\frac{\textstyle \partial u^\pm}{\textstyle \partial x}+d\frac{\textstyle \partial^2 u^\pm}{\textstyle \partial x^2},\quad 0<x<D,\label{Chester1}\\
[6pt]
\displaystyle
u^+(0,t)=u_0(t),\label{Chester2}\\
[6pt]
\displaystyle
u^-(D,t)=u^+(D,t),\label{Chester3}
\end{subnumcases}
with the coefficients
\begin{equation}
\hspace{-0.5cm}
a=a_0,\quad b=\frac{\textstyle \gamma+1}{\textstyle 2},\quad c(x)=\frac{\textstyle C\,a_0 \sqrt{\nu}}{\textstyle R(x)},\quad d=\frac{\textstyle \nu_d}{\textstyle 2}.
\label{CoeffsEDP}
\end{equation}
Menguy-Gilbert's equation (\ref{Chester1}) takes into account nonlinear advection (coefficients $a$ and $b$), viscothermal losses at walls (coefficient $c$) and volumic dissipation (coefficient $d$)  \cite{Chester64,Menguy00}. The operator $\frac{ \partial^{-1/2}}{ \partial t^{-1/2}}$ is the Riemann-Liouville fractional integral of order $1/2$. For a causal function $w(t)$, it is defined by:
\begin{equation}
\begin{array}{lll}
\ds \frac{\textstyle \partial^{-1/2}}{\textstyle \partial t^{-1/2}}w(t)&=& \ds \frac{\textstyle H(t)}{\textstyle \sqrt{\pi\,t}} \ast w,\\
&=&
\ds \frac{\textstyle 1}{\textstyle \sqrt{\pi}}\int_0^t(t-\tau)^{-1/2}\, w(\tau)\,d\tau,
\end{array}
\label{RiemannLiouville}
\end{equation}
where $\ast$ denotes the convolution product, and $H(t)$ is the Heaviside function  \cite{Matignon08}. 

Each wave requires only one boundary condition. The condition for the outgoing wave (\ref{Chester2}) models the acoustic source, linked to the musician. The condition for the incoming wave (\ref{Chester3}) models the Dirichlet condition on pressure at the bell. This condition is the unique coupling between $+$ and $-$ waves. 


\subsubsection{Dispersion analysis}\label{SecResoPhysDisp}

Applying space and time Fourier transforms to (\ref{Chester1}) yields
\begin{equation}
\begin{array}{l}
\ds
i\,d\,k^2\widehat{u^\pm}\pm \left(\left(a - c\chi(\omega)\right)\widehat{u^\pm} \pm \frac{\textstyle b}{\textstyle 2} \widehat {(u^\pm)^2}\right)k\\
[8pt]
\ds 
-\omega\,\widehat{u^\pm} \pm i\, \widehat{\frac{\textstyle a}{\textstyle S}\frac{\textstyle dS}{\textstyle dx}\,u^\pm}=0,
\end{array}
\label{DispersionGuide}
\end{equation}
where the hat refers to the transforms, $k$ is the wavenumber, and $\chi$ is the symbol of the 1/2-integral:
\begin{equation}
\chi(\omega)=\frac{\textstyle 1}{\textstyle \left(i\,\omega\right)^{1/2}}.
\label{ChiDF}
\end{equation}
In the case of constant radius $R$, linear propagation ($b=0$), and no sound diffusivity ($d=0$), the phase velocity $\upsilon=\omega\,/\,\mbox{Re}(k)$ and the attenuation $\alpha=-\mbox{Im}(k)$ of an outgoing wave are deduced explicitly:
\begin{equation}
\begin{array}{l}
\ds
\upsilon=\frac{\textstyle a^2\,\omega-a\,c\sqrt{2\,\omega}+c^2}{\textstyle \displaystyle a\,\omega-c\sqrt{\omega/2}},\\
[8pt]
\ds
\alpha=\frac{\textstyle c}{\textstyle \sqrt{2}}\,\frac{\textstyle \omega^{3/2}}{\textstyle a^2\,\omega-a\,c\,\sqrt{2\,\omega}+c^2}.
\end{array}
\label{CelAttGuide}
\end{equation}
In the case where the viscosity is ignored ($c=d=0$), the phase velocity is equal to $a$, and no attenuation occurs. Otherwise, one has:
\begin{equation}
\begin{array}{llll}
\displaystyle\upsilon(\omega)\mathop{\sim}\limits_{0} - c \sqrt{\frac{2}{\omega}}, &\quad &\quad &\displaystyle\lim_{\omega\rightarrow+\infty}\upsilon(\omega)=a,\\
[8pt]
\displaystyle\alpha(0)=0, &\quad &\quad &\displaystyle\alpha(\omega)\mathop{\sim}\limits_{+\infty} \frac{c}{a^2} \sqrt{\frac{\omega}{2}}.
\end{array}
\label{PropertyGuide}
\end{equation}


\subsection{Mathematical modeling}\label{SecResoMath}

\subsubsection{Diffusive approximation}\label{SecResoMathDiff}

The half-order integral (\ref{RiemannLiouville}) in (\ref{Chester1}) is non-local in time. It requires to keep the memory of the past history of the solution, which is very costly in view of numerical computations. An alternative approach is followed here, based on the diffusive representation of the fractional integral. A change of variables yields \cite{Matignon-These,Diethelm08}
\begin{equation}
\frac{\textstyle \partial^{-1/2}}{\textstyle \partial t^{-1/2}}w(t)=\int_0^{+\infty}\phi(t,\theta)\,d\theta,
\label{I12}
\end{equation}
where the memory variable $\phi$
\begin{equation}
\phi(t,\theta)=\frac{\textstyle 2}{\textstyle \pi}\int_0^t e^{-(t-\tau)\,\theta^2}w(\tau)\,d\tau
\label{Phi12}
\end{equation}
satisfies the ordinary differential equation  
\begin{equation}
\left\{
\begin{array}{l}
\displaystyle
\frac{\partial \phi}{\partial t}=-\theta^2\,\phi+\frac{\textstyle 2}{\textstyle \pi}\,w,\\
[8pt]
\phi(0,\theta)=0.
\end{array}
\right.
\label{ODEI12}
\end{equation}
The diffusive representation  (\ref{I12}) replaces the non-local term (\ref{RiemannLiouville}) by an integral on $\theta$ of functions $\phi(t,\theta)$, which are solutions of local-in-time equations. Integral (\ref{I12}) is then approximated by a quadrature formula
\begin{equation}
\frac{\textstyle \partial^{-1/2}}{\textstyle \partial t^{-1/2}}w(t)\simeq\sum_{\ell=1}^L\mu_{\ell}\,\phi(t,\theta_{\ell})=\sum_{\ell=1}^L\mu_{\ell}\,\phi_{\ell}(t),
\label{RDI12}
\end{equation}
on $L$ quadrature points. Determining the quadrature weights $\mu_{\ell}$ and the nodes $\theta_{\ell}$ is crucial for the efficiency of the diffusive approximation and is discussed further in section \ref{SecResoNumQuad}.


\subsubsection{First-order system}\label{SecResoMathSyst}

The 1/2 integral in  (\ref{Chester1}) is replaced by its diffusive approximation (\ref{RDI12}) and by the set of differential equations satisfied by the memory functions $\phi_{\ell}$ (\ref{ODEI12}). It follows the two systems for $+$ and $-$ waves
\begin{subnumcases}{\label{SystComplet}}
\displaystyle
\frac{\textstyle \partial u^\pm}{\textstyle \partial t} + \frac{\textstyle \partial}{\textstyle \partial x}\left(\pm a \textstyle u^\pm+b\,\displaystyle \frac{\textstyle (u^\pm)^2}{\textstyle 2}\right)\pm\frac{\textstyle a}{\textstyle S}\,\frac{\textstyle dS}{\textstyle dx}\,u^\pm\\
\displaystyle \hspace{0.3cm}
=\pm c\sum_{\ell=1}^L\mu_{\ell}\phi_{\ell}+d\,\frac{\textstyle \partial^2 u^\pm}{\textstyle\partial x^2},\quad 0<x<D,\label{SystComplet1}\\
[6pt]
\displaystyle
\frac{\textstyle \partial \phi^\pm_{\ell}}{\textstyle \partial t}-\frac{\textstyle 2}{\textstyle \pi}\,\frac{\textstyle \partial u^\pm}{\textstyle \partial x}=-\theta_{\ell}^2\,\phi^\pm_{\ell},\,\ell=1,\dots,L,\label{SystComplet2}\\
[6pt]
\displaystyle
u^+(0,t)=u_0(t),\label{SystComplet3}\\
[6pt]
\displaystyle
u^-(D,t)=u^+(D,t),\label{SystComplet4}\\
[6pt]
\ds u(x,0)=v(x),\,\phi_\ell(x,0)=0, \,\ell=1,\dots,L.\label{SystComplet5}
\end{subnumcases}
The $(L+1)$ unknowns are gathered in the vectors ${\bf U}^\pm$: 
\begin{equation}
{\bf U}^\pm=\left(u^\pm,\phi^\pm_1,\cdots,\,\phi^\pm_L\right)^T.
\label{VecU}
\end{equation}
Then the systems (\ref{SystComplet}) are recast as:
\begin{equation}
\hspace{-0.3cm}
\frac{\textstyle \partial}{\textstyle \partial t}{\bf U}^\pm+\frac{\textstyle \partial}{\textstyle \partial x}{\bf F}^\pm({\bf U}^\pm)={\bf S}^\pm\,{\bf U}^\pm+{\bf G}\,\frac{\textstyle \partial^2}{\textstyle \partial x^2}{\bf U}^\pm,
\label{SystHyper}
\end{equation}
where ${\bf F}^\pm$ are the nonlinear flux functions 
\begin{equation}
\hspace{-0.8cm}
{\bf F}^\pm({\bf U}^\pm)=\left(\pm au^\pm + b\frac{\textstyle (u^\pm)^2}{\textstyle 2},-\frac{\textstyle 2}{\textstyle \pi}u^\pm,\cdots,-\frac{\textstyle 2}{\textstyle \pi}u^\pm\right)^T
\label{Fnonlin}
\end{equation}
and ${\bf G}$ is the $(L+1)\times (L+1)$ diagonal matrix with terms $\mbox{diag}(d,\,0,\cdots,\,0)$. The relaxation matrices ${\bf S}^\pm$ include both a geometrical term, due to the variation of section, and physical terms, related to the diffusive approximation of viscothermal losses: 
\begin{equation}
{\bf S}^\pm=
\left(
\begin{array}{cccc}
\displaystyle
\mp \frac{\textstyle a}{\textstyle S}\,\frac{\textstyle dS}{\textstyle dx} & \pm c\,\mu_1 & \cdots & \pm c\,\mu_L\\
0 & -\theta_1^2 & & \\
\vdots &  & \ddots & \\
0 & & & -\theta_L^2 
\end{array}
\right).
\label{MatS}
\end{equation}


\subsubsection{Properties}\label{SecResoMathProp}

In brass musical instruments, the volumic losses are negligible compared to the viscothermal losses  \cite{Sugimoto91,Menguy00}. Consequently, the dynamics of systems (\ref{SystHyper}) is essentially unchanged when taking ${\bf G}={\bf 0}$. In this case, properties of the solutions rely on the Jacobian matrices ${\bf J}^\pm=\frac{\partial {\bf F}^\pm}{\partial {\bf U}^\pm}$. Some properties are listed here without proof; interested readers are referred to standard textbooks for more details about hyperbolic systems  \cite{LeVeque92,Godlewski96}.

The eigenvalues $\lambda_j^\pm$ of ${\bf J}^\pm$ are real ($j=1,\dots,L+1$):
\begin{equation}
\lambda_1^\pm=\pm a+b\,u^\pm,\quad \lambda_j^\pm=0;
\end{equation}
Assuming $\lambda_1^\pm \neq 0$, the matrices of eigenvectors ${\bf R}^\pm=({\bf r}^\pm_1|{\bf r}^\pm_2|...|{\bf r}^\pm_{L+1})$ are
\begin{equation}
{\bf R}^\pm=
\left(
\begin{array}{cccc}
1 & 0 & \cdots & 0\\
[6pt]
\ds  \frac{-2}{\pi\left(\mp a+bu^\pm\right)} & 1 & \\
\vdots & & \ddots & \\
\ds  \frac{-2}{\pi\left(\mp a+bu^\pm\right)} & & & 1
\end{array}
\right),
\label{MatR}
\end{equation}
and they are invertible if $u^\pm \neq \pm a/b$, which is consistent with the assumption of weak nonlinearity: the systems (\ref{SystHyper}) are hyperbolic, but not strictly hyperbolic (multiple eigenvalues). The characteristic fields satisfy:
\begin{equation}
\hspace{-0.5cm}
{\bf \nabla \lambda_1}.{\bf r}_1^\pm=b\neq 0, \quad {\bf \nabla \lambda_j}.{\bf r}_j^\pm=0, \hspace{0.2cm} j=1,\dots,L+1,
\end{equation}
where the gradient is calculated with respect to each coordinate of $\bf U$ in (\ref{VecU}), as defined in   \cite[p77]{Toro99}. Consequently, there exists 1 genuinely nonlinear wave (shock wave or rarefaction wave), and $L$ linearly degenerate waves (contact discontinuities).

Moreover, the eigenvalues of the relaxation matrices ${\bf S}^\pm$ are ($j=1,\dots,L+1$):
\begin{equation}
\kappa_1=\mp\frac{\textstyle a}{\textstyle S}\,\frac{\textstyle dS}{\textstyle dx}, \hspace{0.2cm} \kappa_j=-\theta_j^2.
\end{equation}
Assuming that the quadrature nodes are sorted by increasing order ($\theta_1<\theta_2<\cdots<\theta_L$), the spectral radius of ${\bf S}^\pm$ is
\begin{equation}
\varrho({\bf S}^\pm)=\max\left(\max_{x\in[0,D]}\frac{a}{\textstyle S}\,\frac{\textstyle dS}{\textstyle dx}, \theta_L^2\right).
\label{RayonSpectral}
\end{equation}
This quantity becomes large in the case of a rapidly-varying section, or with a large maximal quadrature node of the fractional integral (see the next section).

Lastly, a Fourier analysis of (\ref{SystComplet}) leads to a similar dispersion relation than (\ref{DispersionGuide}). The symbol $\chi$ in (\ref{ChiDF}) has only to be replaced by the symbol of the diffusive approximation 
\begin{equation}
\tilde{{\chi}}(\omega)=\frac{\textstyle 2}{\textstyle \pi}\sum_{\ell=1}^L\frac{\textstyle \mu_{\ell}}{\textstyle \theta_{\ell}^2+i\,\omega}.
\label{ChiAD}
\end{equation}


\subsection{Numerical modeling}\label{SecResoNum}

\subsubsection{Quadrature methods}\label{SecResoNumQuad}
 
Basically, two strategies exist to determine the quadrature weights $\mu_{\ell}$ and nodes $\theta_{\ell}$ in (\ref{RDI12}), which are involved in the relaxation matrices (\ref{MatS}). The first strategy relies on Gaussian polynomials, for instance the modified Gauss-Jacobi polynomials  \cite{Birk10,NRPAS}. This approach offers a solid mathematical framework, but very low convergence rate is obtained  \cite{Lombard14}. As a consequence, a large number $L$ of memory variables is required to describe the fractional integral by a quadrature formula (\ref{RDI12}), which penalizes the computational efficiency.

An alternative approach is followed here, based on an optimization procedure on the symbols (\ref{ChiDF}) and (\ref{ChiAD}). Given a number $K$ of angular frequencies $\omega_k$, the following cost function is introduced:  
\begin{equation}
\begin{array}{l}
{\cal J}\left({\{(\mu_\ell,\theta_\ell)\}}_\ell\,;L,K\right)=
\displaystyle \sum_{k=1}^{K}\left|\frac{\textstyle {\tilde \chi}(\omega_k)}{\textstyle \chi(\omega_k)}-1\right|^2,\\
[8pt]
\hspace{1cm}
=
\ds \sum_{k=1}^{K}\left|\frac{\textstyle 2}{\textstyle \pi} \sum_{\ell=1}^{L}\mu_\ell\,\frac{\textstyle (i\omega_k)^{1/2}}{\theta_{\ell}^2+i\omega_k}-1\right|^2,
\end{array}
\label{Objective}
\end{equation}
to be minimized with respect to the parameters  $\{(\mu_\ell,\theta_\ell)\}_\ell$ for $\ell=1,\dots,L$. A nonlinear optimization with a positivity constraint $\mu_{\ell}\geq 0$ and $\theta_{\ell}\geq 0$ is adopted. For this purpose, one implements the SolvOpt algorithm,  \cite{Rekik11,Blanc13} based on Shor's iterative method  \cite{Shor85}. Initial values in the optimization algorithm have to be chosen with care. This is done by using coefficients obtained by the modified orthogonal Jacobi polynomials  \cite{Birk10,Lombard14}. Finally, the angular frequencies $\omega_k$ in (\ref{Objective}) are linearly spaced on a logarithmic scale on the optimization range $[\omega_{\min},\omega_{\max}]$:
\begin{equation}
\omega_k = \omega_{\min}\left( \frac{\omega_{\max}}{\omega_{\min}}\right)^{\!\frac{k-1}{K-1}},\hspace{0.5cm}k=1,\cdots,L.
\label{OmegaK}
\end{equation}
The number $K$ of angular frequencies $\omega_k$ is chosen equal to the number $L$ of diffusive variables.


\subsubsection{Numerical scheme}\label{SecResoNumSchem}

To perform numerical integration of systems (\ref{SystHyper}), a uniform grid for space is introduced with step $\Delta x=D/N_x$, as well as a variable time step $\Delta t_n$ (denoted $\Delta t$ for the sake of simplicity). The approximation of the exact solution ${\bf U}^\pm(x_j = j\,\Delta x, t_n = t_{n-1}+\,\Delta t)$ is denoted ${\bf U}_j^{n\pm}$. A direct explicit discretization of (\ref{SystHyper}) is not optimal, since the numerical stability requires  \cite{These-Blanc}
\begin{equation}
\Delta t\leq \min\left(\frac{\Delta x}{a_{\max}^n},\frac{2}{\varrho({\bf S}^\pm)}\right),
\label{CFLdirect}
\end{equation}  
where 
  $a_{\max}^n = \max |\pm a + b u_j^{(n)\pm} |$ is the maximum numerical velocity at time $t_n$. As shown in (\ref{RayonSpectral}), 
the spectral radius of the relaxation matrices ${\bf S}^\pm$ grows with the maximal node of quadrature, which penalizes the standard Courant-Friedrichs-Lewy (CFL) condition of stability (\ref{CFLdirect}).

A more efficient strategy is adopted here. Equations (\ref{SystHyper}) are split into a propagative step
\begin{equation}
\frac{\textstyle \partial}{\textstyle \partial t}{\bf U}^\pm+\frac{\textstyle \partial}{\textstyle \partial x}{\bf F}({\bf U}^\pm)={\bf G}\,\frac{\textstyle \partial^2}{\textstyle \partial x^2}{\bf U}^\pm,
\label{SplitPropa}
\end{equation}
and a relaxation step
\begin{equation}
\frac{\partial}{\partial t}{\bf U}^\pm={\bf S}^\pm\,{\bf U}^\pm.
\label{SplitDiffu}
\end{equation}
The discrete operators associated to (\ref{SplitPropa}) and (\ref{SplitDiffu}) are denoted by ${\bf H}^\pm_a$ and ${\bf H}^\pm_b$, respectively. Strang splitting is then used to do a step forward from $t_n$ to $t_{n+1}$ by solving successively (\ref{SplitPropa}) and (\ref{SplitDiffu}) with adequate time steps:  \cite{LeVeque92}

\begin{equation}
\begin{array}{lll}
\ds {\bf U}_{j}^{(1)\pm} &= &{\bf H}^\pm_{b}\left(\frac{\Delta t}{2}\right)\,{\bf U}_{j}^{(n)\pm},\\
[8pt]
\ds {\bf U}_{j}^{(2)\pm} &= &{\bf H}^\pm_{a}\left(\Delta t\right)\,{\bf U}_{j}^{(1)\pm},\\
[8pt]
\ds {\bf U}_{j}^{(n+1)\pm} &= &{\bf H}^\pm_{b}\left(\frac{\Delta t}{2}\right)\,{\bf U}_{j}^{(2)\pm}.
\end{array}
\label{AlgoSplitting}
\end{equation}
Equation (\ref{SplitPropa}) corresponding to the propagative part is solved with a second-order TVD scheme (Total Variation Diminishing) for hyperbolic equations:  \cite{Godlewski96}
\begin{equation}
\begin{array}{lll}
\ds
{\bf U}_i^{(n+1)\pm} &=& \ds {\bf U}_i^{(n)\pm} -\frac{\Delta t}{\Delta x} ({\bf F}_{i+1/2}^{\pm}- {\bf F}_{i-1/2}^{\pm})\\
[8pt]
\ds 
&+&\ds {\bf G} \frac{\Delta t}{\Delta x²}( {\bf U}_{i+1}^{(n)\pm}-2 {\bf U}_i^{(n)\pm}+ {\bf U}_{i-1}^{(n)\pm}), 
\end{array}
\label{TVD}
\end{equation}
where ${\bf F}^\pm_{i\pm1/2}$ is the numerical flux function for Burgers equation in (\ref{Fnonlin}), 
and $\ell=1,\dots,L$. Defining the discrete P\'eclet number Pe
\begin{equation}
\mbox{Pe}=a_{\max}^{n}\,\frac{\textstyle \Delta x}{\textstyle 2\,d}\gg 1,
\label{Peclet}
\end{equation}
the discrete operator ${\bf H}^\pm_{a}$ in (\ref{TVD}) is stable if  \cite{Sousa03}
\begin{equation}
\varepsilon=\frac{a_{\max}^{n}\,\Delta t}{\Delta x}\leq \left(1+\frac{\textstyle 1}{\textstyle \mbox{Pe}}\right)^{-1} \approx 1-\frac{\textstyle 1}{\textstyle \mbox{Pe}} \approx 1,
\label{CFL}
\end{equation}
which recovers the optimal CFL condition. Therefore, thanks to the splitting strategy, the value of $\Delta t$ is no more penalized by the spectral radius of {\bf S$^\pm$}. 

Equation (\ref{SplitDiffu}) of the diffusive part is solved exactly:
\begin{equation}
{\bf H}^\pm_b\left({\textstyle\frac{\Delta t}{2}}\right)\,{\bf U}^\pm_j = \exp\left({\textstyle{\bf S}^\pm\,\frac{\Delta t}{2}}\right)\,{\bf U}^\pm_j,
\label{SplitDiffuExp}
\end{equation}
with the matrix exponential deduced from (\ref{MatS})
\begin{equation}
\begin{array}{l}
\hspace{-0.8cm}
e^{{\bf S}^\pm \tau}=\\
\ds
\hspace{-1cm}
\left(
\begin{array}{cccc}
\displaystyle
e^{-\Omega^\pm \tau} & \delta_1^\pm\left(e^{-\Omega^\pm \tau}-e^{-\theta_1^2 \tau}\right) & \cdots & \delta_L^\pm\left(e^{-\Omega^\pm \tau}-e^{-\theta_L^2 \tau}\right)\\
0 & e^{-\theta_1^2 \tau} & & \\
\vdots &  & \ddots & \\
0 & & & e^{-\theta_L^2 \tau} 
\end{array}
\right),
\end{array}
\label{MatExpS}
\end{equation}
and the coefficients  ($\ell=1,\dots,L$)
\begin{equation}
\Omega^\pm=\pm \frac{a}{S}\frac{dS}{dx},\hspace{0.5cm}
\delta_\ell^\pm=\pm \frac{\textstyle c\,\mu_\ell}{\textstyle \theta_\ell^2 - \Omega^\pm}.
\end{equation}
This relaxation step is unconditionally stable. Without viscothermal losses ($c=0$), the matrix exponential (\ref{MatExpS}) degenerates towards the scalar $e^{-\Omega^\pm \tau}$. The physically-realistic case $\frac{dS}{dx}>0$ yields a decreasing amplitude of $u^+$ as $x$ increases. Inversely, it yields an increasing amplitude of $u^-$ as $x$ decreases.

The Jacobian matrices and the relaxation matrices do not commute: ${\bf J}^\pm{\bf S}^\pm \neq {\bf S}^\pm{\bf J}^\pm$. Consequently, the splitting (\ref{AlgoSplitting}) is second-order accurate  \cite{LeVeque92}. It is stable under the CFL condition (\ref{CFL}).


\subsection{Validation}\label{SecResoVal}

\subsubsection{Configuration}\label{SecResoValConf}

\begin{table}[htbp]
\begin{center}
{\renewcommand{\arraystretch}{1.2}
\renewcommand{\tabcolsep}{0.2cm}
\begin{tabular}{cccccc}	
\hline
$\gamma$ & $p_0$ (Pa) & $\rho_0$ (kg/m$^{3}$) & Pr      & $\nu$ (m$^2$/s)     & $\mu_v/\mu$ \\
\hline  
1.403    & $ 10^5$    & 1.177                 & 0.708   & $1.57\cdot 10^{-5}$ & 0.60        \\
\hline
\end{tabular}}
\end{center}
\caption{\label{TabParam}Physical parameters of air at $15\,^{\circ}\mathrm{C}$.}
\end{table}

In all the tests, one considers a circular tube of length $D=1.4$~m, with an entry radius $R(0)=7$~mm. The physical parameters given in table \ref{TabParam} are used to determine the coefficients (\ref{CoeffsEDP}) in (\ref{SystComplet1}). 
The tube is discretized on $N_x$ points in space, the value of $N_x$ being precised for each test case. 
At each iteration, the time step $\Delta t$ is deduced from the condition (\ref{CFL}), where the CFL number is $\varepsilon=0.95$. 


\subsubsection{Diffusive approximation}\label{SecResoValDA}

\begin{figure}[htbp]
\begin{center}
\begin{tabular}{c}
(i)\\
\includegraphics[scale=1.2]{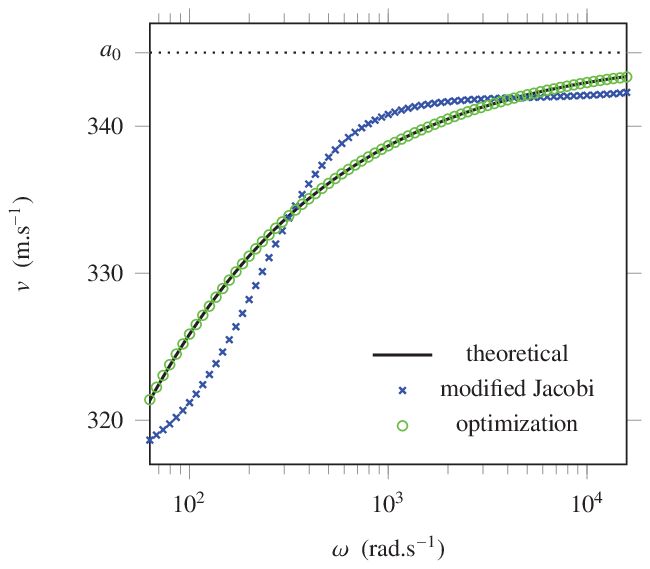} \\
(ii)\\
\includegraphics[scale=1.2]{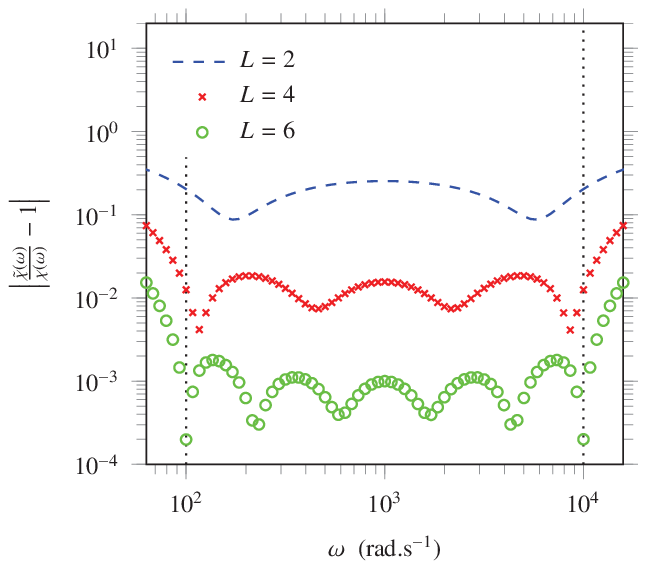} 
\end{tabular}
\caption{\label{FigDispersionOpti} Approximation of the fractional derivative. (i) Phase velocity of the Menguy-Gilbert model (\ref{Chester}) and of the diffusive model (\ref{SystComplet}). The horizontal dotted line denotes the sound velocity $a_0$. (ii) Error $\left|\, {\tilde \chi (\omega)}/{\chi(\omega)}-1\right|$. Vertical dotted lines show limits of the range $[\omega_{\min},\omega_{\max}]$ where the diffusive approximation is optimized.} 
\end{center}
\end{figure}

The first test investigates the accuracy of the diffusive approximation to model fractional viscothermal losses (section \ref{SecResoNumQuad}). The nonlinearity and the volumic attenuation are neglected ($b=0$, $d=0$), and the radius $R$ is constant.  The tube is discretized on $N_x=200$ points in space. Based on the discussion of section \ref{SecResoNumQuad}, comparison is performed between a modified Gauss-Jacobi quadrature and an optimized quadrature. The reference solution is the phase velocity of the linear Menguy-Gilbert model (\ref{CelAttGuide}), where the symbol $\chi$ is given by (\ref{ChiDF}). Conversely, the phase velocity of the diffusive model relies on the symbol (\ref{ChiAD}).

Figure~\ref{FigDispersionOpti}-(i) compares these different phase velocities, using $L=6$ memory variables. Large errors are obtained when the modified Gauss-Jacobi quadrature is used. On the contrary, the agreement between exact and approximate phase velocities is far better when the optimization with constraint is used. In this latter case, the optimization range $\left[\omega_{\min},\omega_{\max}\right]$ is set to $[10^2,10^4]$~rad/s. 

From now on, optimization with constraint is chosen. To see more clearly the error induced by the optimization (\ref{Objective}), figure~\ref{FigDispersionOpti}-(ii) displays the error of modeling $\left|\, {\tilde \chi (\omega)}/{\chi(\omega)}-1\right|$ on a logarithmic scale. The optimization of the coefficients $(\mu_\ell,\theta_\ell)$ is performed with different numbers of memory variables $L$. The error decreases approximately by a factor 10 when $L$ is doubled. In the following numerical experiments, the viscothermal losses have been accounted for by $L=6$ memory variables, optimized over the range of frequency $\left[\omega_{\min},\omega_{\max}\right] = [10^2,10^4]$~rad/s.
 

\subsubsection{Nonlinear propagation}\label{SecResoValPropaNL}

\begin{figure}[htbp]
\begin{center}
\begin{tabular}{c}
(i)\\
\includegraphics[scale=1.3]{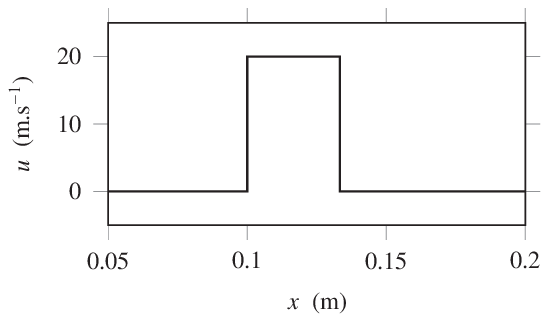} \\
(ii)\\
\includegraphics[scale=1.3]{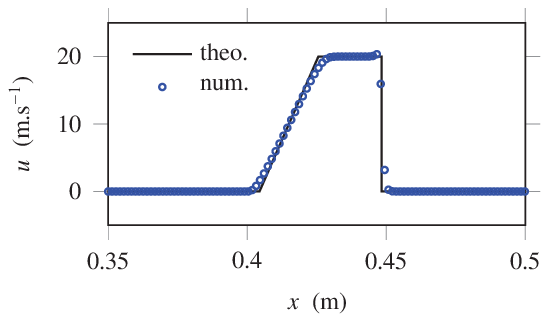} \\
(iii)\\
\includegraphics[scale=1.3]{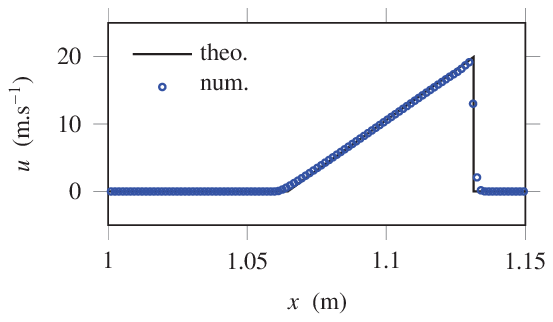}
\end{tabular}
\caption{\label{FigPorteNL} Nonlinear wave propagation. (i) Initial data $v(x)$. (ii) comparison between exact and numerical values of the outgoing velocity at $t\approx 0.88$~ms. (iii) idem at $t\approx 2.8$~ms.}
\end{center}
\end{figure}

The second test concerns the modeling of nonlinear waves by the TVD scheme (\ref{TVD}). For this purpose, losses are neglected ($c=0$, $d=0$), and the radius of the tube is constant. The forcing in (\ref{SystComplet3}) is null. The initial data (\ref{SystComplet5}) is a rectangular pulse with a $20$~m/s amplitude and a wavelength $\lambda=0.03$~m.  The tube is discretized on $N_x=1000$ points in space. 

Figure~\ref{FigPorteNL} displays the numerical solution and the exact solution at various instants. The latter is derived from the elementary solutions to the Riemann problem  \cite{LeVeque92}. In (ii), one observes a outgoing shock wave followed by a rarefaction wave. In (iii), the rarefaction has reached the shock. In each case, agreement is observed between numerics and analytics, despite the non-smoothness of the solution. In particular, the shock propagates at the good speed, which reveals that the Rankine-Hugoniot condition is correctly taken into account.


\subsubsection{Linear propagation with a varying cross section area}\label{SecResoValVar}

\begin{figure}[htbp]
\begin{center}
\begin{tabular}{c}
(i)\\
\includegraphics[scale=1.4]{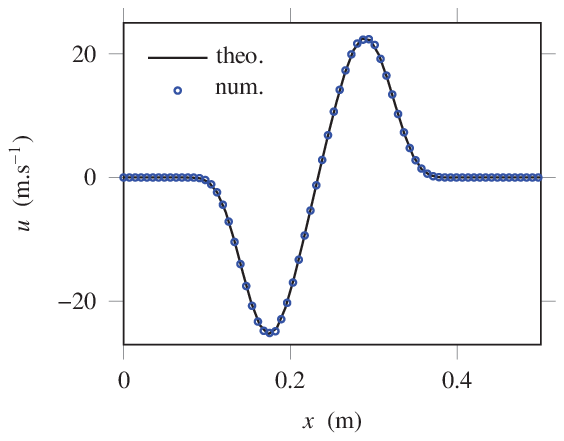} \\
(ii)\\
\includegraphics[scale=1.4]{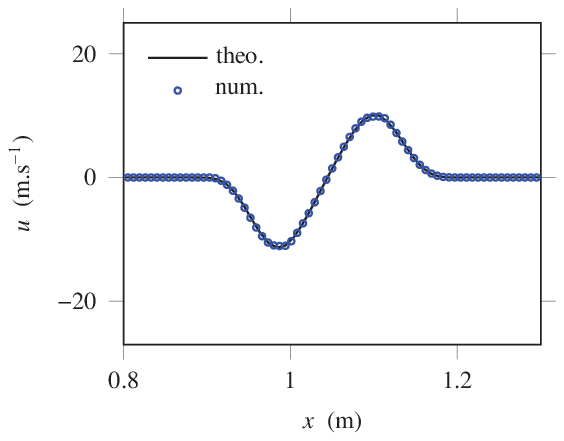}
\end{tabular}
\caption{\label{FigSecVarLin} Tube with exponentially-varying cross section area (\ref{SectionVar}). Snapshots of the exact and numerical velocity of the outgoing wave, at $t\approx 1.2$~ms (i) and $t\approx 3.5$~ms (ii).}
\end{center}
\end{figure}

The third test focuses on a variable cross section area$S(x)$, with a radius varying exponentially from $R(0)=7$~mm to $R(D)=2\,R(0)$:
\begin{equation}
\displaystyle S(x)=\pi\,\left(R(0)\,2^{x/D}\right)^2, \quad 0\leq x\leq D.
\label{SectionVar}
\end{equation}
The tube is discretized on $N_x=200$ points in space. Linear propagation is assumed ($b=0$), and the dissipation effects are neglected ($c=0$, $d=0$). Only the outgoing wave is considered. The discretization of the variable radius involves the relaxation parts of the splitting (\ref{AlgoSplitting}): only the component $e^{-\Omega^+\tau}$ in (\ref{MatExpS}) is non-null. The exciting source in (\ref{SystComplet3}) is a smooth  combination of truncated sinusoidal wavelets:
\begin{equation}
\hspace{-0.4cm}
u_0(t)=\left\{
\begin{array}{ll}
\displaystyle V\sum_{m=1}^4 a_m \sin\,(b_m\,\omega_c\,t) &\mbox{if }\, 0\leq t\leq {\textstyle \frac{\displaystyle 1}{\displaystyle f_c}},\\
0 &\mbox{else},
\end{array}
\right. 
\label{Source}
\end{equation}
with amplitude $V=20$~m/s, central frequency $f_c={\omega_c}/{2\,\pi}=1$~kHz and coefficients $b_m=2^{m-1}$, $a_1=1$, $a_2=-21/32$, $a_3=63/768$ and $a_4=-1/512$. The exact solution is straightforwardly deduced from the method of characteristics 
\begin{equation}
u^+(x,t)=\exp\left(-\Omega^+\frac{x}{a}\right)\,u_0\left(t-\frac{x}{a}\right).
\label{ExactSecVar}
\end{equation}
Figure~\ref{FigSecVarLin} displays a snapshot of the velocity $u^+$, at two successive instants. Agreement between numerical and theoretical results is obtained. As deduced from (\ref{ExactSecVar}), the amplitude of the wave decreases as the wavefront advances.


\subsubsection{Simulation of an input impedance}\label{SecResoValImp}

\begin{figure}[h!]
\begin{center}
\includegraphics[scale=1.2]{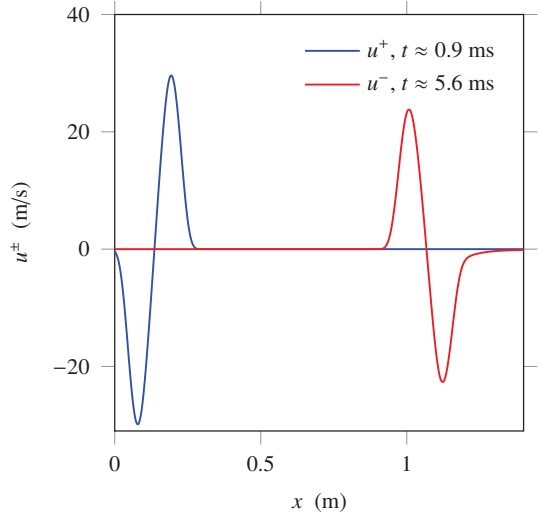}
\caption{Snapshot of the outgoing and ingoing velocity at to different times, where propagation is linear and the tube has a constant cross section area.\label{FigPropaAR} 
}
\end{center}
\end{figure}

\begin{figure}[h!]
\begin{center}
\begin{tabular}{c}
(i)\\
\includegraphics[scale=0.99]{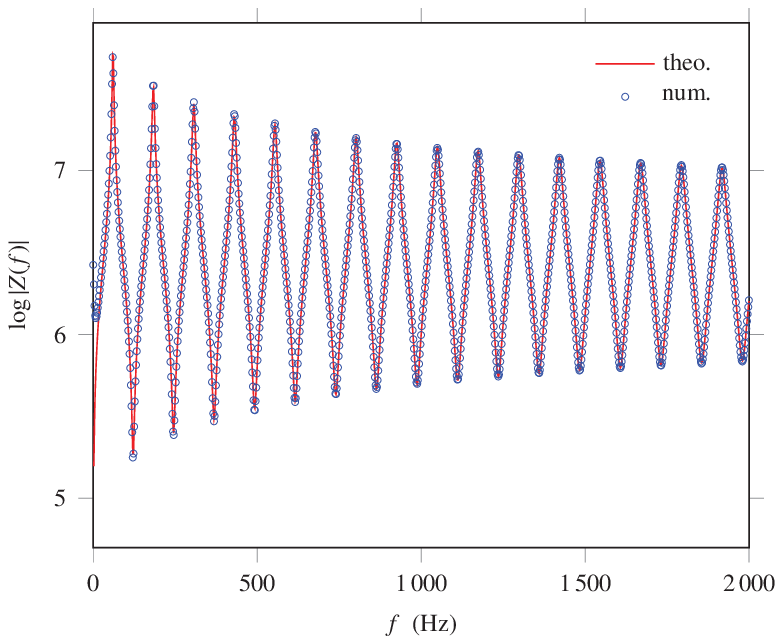} \\
(ii)\\
\includegraphics[scale=0.99]{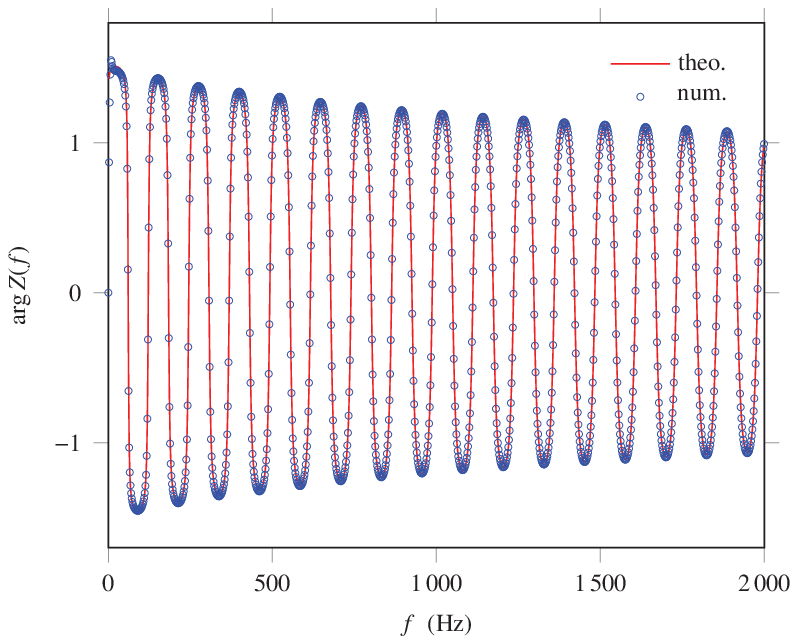}
\end{tabular}
\caption{Input impedance $Z$: modulus (i) and phase (ii). Comparison between simulated and exact values. \label{FigImpedance} }
\end{center}
\end{figure}

The coupling between the advection and the diffusive approximation of the viscothermal losses is studied here, as well as the interaction between the simple waves (\ref{SystComplet4}). A constant radius $R=7$~mm is considered. Linear wave propagation is assumed ($b=0$). The tube is discretized using $N_x=1000$ points in space. The exciting source in (\ref{SystComplet3}) is the wavelet (\ref{Source}), with the same central frequency $f_c$ and amplitude $V$ as in section \ref{SecResoValVar}.

When the outgoing wave "+" reaches the limit of the tube ($x=D$), the incoming wave "$-$" is generated and propagates along the decreasing $x$. The velocity is displayed at two different times on figure \ref{FigPropaAR}. Due to viscothermal losses, the amplitude of the wave diminishes during the simulation. Also, after 5.6~ms of propagation, the waveform is not symmetric anymore, which illustrates the dispersive nature of the propagation.

Now, we take $N_x=2000$ to compute the input impedance. This high number of discretization points is required to get a high frequency resolution to calculate the input impedance. A receiver at $x=0$ records the pressure $p^-(0,t_n)$. The outgoing pressure $p^+(0,t_n)$ is known, corresponding to the exciting source. Fourier transforms in time of $p^\pm$ yield an estimate of the input impedance $Z$ of the tube:
\begin{equation}
Z(\omega)=Z_c\,\frac{\textstyle 1+r(\omega)}{\textstyle 1-r(\omega)},
\label{Impedance}
\end{equation}
with
\begin{equation}
Z_c=\frac{\textstyle \rho_0\,a_0}{\textstyle S},\qquad r(\omega)=\frac{\textstyle {\widehat{p^-}}(0,\omega)}{\textstyle {\widehat{p^+}}(0,\omega)}.
\label{Ratio}
\end{equation}
Figure~\ref{FigImpedance} shows the modulus and the phase of the input impedance deduced from the numerical simulations. These quantities are compared to their analytical approximation given by  \cite{Chaigne08}
\begin{equation}
\hspace{-0.5cm}
Z=i\,Z_c\tan(kD),\quad k=\frac{\textstyle \omega}{\textstyle a_0}-i\,(1+i)\,3\cdot 10^{-5}\frac{\textstyle \sqrt{f}}{\textstyle R}.
\label{Ztheo}
\end{equation}
Excellent agreement is observed, except around null frequency. These small differences are due to the spectrum of the wavelet (\ref{Source}), which vanishes when $f=0$ Hz. It results numerical inaccuracies in the ratio (\ref{Ratio}).


\subsubsection{Complete Menguy-Gilbert model }\label{SecResoValTotal}

\begin{figure}[htbp]
\begin{center}
\includegraphics{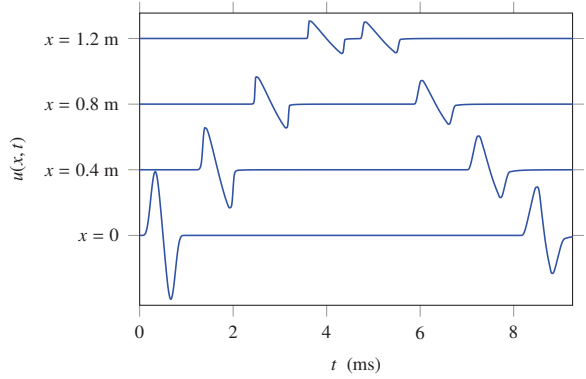} 
\caption{\label{FigSismo} Complete model of resonator: nonlinear wave propagation, viscothermal losses, volumic dissipation, variable section. Time-history of the velocity at four receivers along the exponential horn.}
\end{center}
\end{figure}

As a last experiment, we take into account all the effects in (\ref{Chester}). The exciting source is (\ref{Source}). The tube is discretized using $N_x=300$ points in space. 
The pressure is recorded at four receivers uniformly distributed along the exponential horn.

Figure~\ref{FigSismo} displays the time history of the velocity $u=u^++u^-$ at these receivers. In each case, the velocity $u^+$ of the outgoing wave is recorded first (up to $t\approx 5$~ms), followed by the velocity $u^-$ of the incoming wave reflected by the end of the cylinder. As $x$ increases, the amplitude of $u^+$ decreases. It is due to three factors: the emergence of shocks, the intrinsic losses, and the increase of the cross section area (see section \ref{SecResoValVar}). 

On the contrary, the amplitude of $u^-$ increases as the location of receivers decreases, from $x=1.2$ m downto $x=0$ m. This perharps counter-intuitive observation is explained as follows: in the direction of propagation of the incoming wave (decreasing $x$), the cross section of the guide decreases. It results in an increasing amplitude, exceeding the effect of the losses and of the nonlinearity.

Moreover, at each receiver, $u^-$ appears to be less distorted than $u^+$. Indeed, due to the boundary condition (\ref{SystComplet4}), the incoming wave $u^-$ experiences nonlinear effects which balance the ones experienced by the outgoing wave $u^+$ (see (\ref{SystComplet1})). The balance is complete in a lossless description and if a shock does not occur on  $u^+$. Here, a closer view on the figure would reveal that a shock indeed occur before $x=1.2$~m.

Lastly, at each receiver, one notices that $u^-$ has a smaller amplitude than $u^+$. This is due to two causes. First of all, the losses act both on $u^+$ and $u^-$, hence the effects are cumulative. Secondly, a shock is a dissipative phenomenon and $u^-$ would have a smaller amplitude than $u^+$ if a shock occurs on $u^+$ (which is the case here), even if visco-thermal and volumic losses were ignored.


\section{Exciter}\label{SecExc}

\subsection{Physical modeling}\label{SecExcPhys}

\begin{figure}[htbp]
\begin{center}
\includegraphics[scale=1.2]{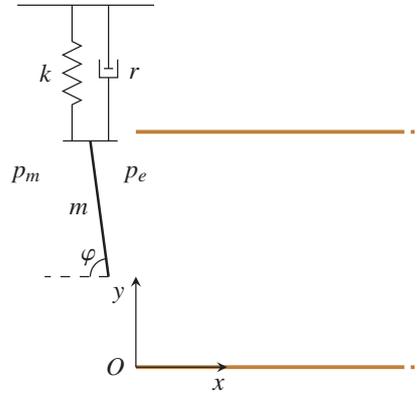}
\caption{\label{FigLevres} One-mass model for the lips.}
\end{center}
\end{figure}

The musician's lips are modeled by a one-mass mechanical oscillator at the entry of the resonator  \cite{Elliott82}. Only the vertical displacement of the top lip is modeled; the interaction with the static bottom lip is ignored. The top lip is modeled by a thin rigid rectangular plate of height $h$ and width $l$. It makes an angle $\varphi$ with the horizontal $x$-axis, so that the projected surface of the lip on the vertical axis is
\begin{equation}
\displaystyle
A=h\,l\sin\varphi.
\label{SurfProj}
\end{equation}
A spring with stiffness $k$ and a damper with coefficient $r$ are put over the lip of mass $m$ (figure~\ref{FigLevres}). The pressure in the musician's mouth is $p_m(t)$; the acoustical pressure $p_e$ at the entry of the resonator ($x=0$) depends upon the opening $y$ of the lips and upon time $t$:
\begin{equation}
\begin{array}{lll}
p_e(y,t)&=& p_e^+(y,t)+p_e^-(y,t),\\
[8pt]
&=& p^+(0,t)+p^-(0,t),\\
\end{array}
\label{Pe}
\end{equation}
The balance of forces yields the ordinary differential equation satisfied by $y$:
\begin{subnumcases}{\label{OscMeca}}
\displaystyle
m\ddot{y}+r\dot{y}+k(y-y_\textit{eq})=f(y,t),\label{OscMeca1}\\
[6pt]
\displaystyle
y(0)=y_0,\hspace{0.5cm}\dot{y}(0)=y_1,\label{OscMeca3}
\end{subnumcases}
where $y_\textit{eq}$ is the equilibrium position of the free oscillator, 
\begin{equation}
f(y,t)=A\,(p_m(t)-p_e(y,t))\label{AeroForce}
\end{equation}
is the aeroacoustic force applied to the lip, and $(y_0,y_1)$ are the initial conditions.

The flow is assumed to be stationary, incompressible, laminar and inviscid in the musician's mouth and under the lip. Consequently, Bernoulli's equation and the conservation of mass can be applied. The sudden cross section variation behind the lip creates a turbulent jet which dissipates all its kinetic energy without pressure recovery in the mouthpiece  \cite{hirschberg95}. It follows  \cite{McIntyre83, These_Vergez}
\begin{equation}
\begin{array}{l}
\ds
\hspace{-1cm}
p_e(y,t) = \\
[8pt]
\hspace{-1cm}
\left\{\begin{array}{l}
\displaystyle
2 p^-_e -\frac{\xi}{2}\psi y\left(\psi y-\sqrt{\psi^2 y^2 + 4\left| p_m - 2 p^-_e \right|} \right) \mbox{if } y > 0,\\
\\
\displaystyle
2 p^-_e \mbox{ else.}
\end{array}\right.
\end{array}
\label{ExciterDepNL}
\end{equation}
The coefficients in (\ref{ExciterDepNL}) are
\begin{equation}
\hspace{-0.8cm}
\xi(y,t)=\mbox{sgn}(p_m(t)-p_e(y,t))=\mbox{sgn}(p_m(t)-2p^-_e(y,t))
\label{CoeffsCoupl1}
\end{equation}
and
\begin{equation}
\psi = l\,Z_c \, \sqrt{\frac{2}{\rho_0}}=l\sqrt{2\,\rho_0}\frac{a_0}{S(0)}.
\label{CoeffsCoupl2}
\end{equation}


\subsection{Numerical modeling}\label{SecExcNum}

\subsubsection{Numerical scheme}\label{SecExcNumScheme}

The numerical integration of (\ref{OscMeca}) relies on a variable time step $\Delta t_n$, noted  $\Delta t$ for sake of simplicity; as shown further in section \ref{SecExpAlgo}, it is the time step used for wave propagation in the resonator. The approximation of the exact solution $y(t_n)$ is denoted $y_n$. Similarly, $\dot{y}(t_n)$ and $\ddot{y}(t_n)$ are approximated by $\dot{y}_n$ and $\ddot{y}_n$, respectively. 

The Newmark method is applied to (\ref{OscMeca}). This method, which relies upon two coefficients $\beta$ and $\eta$, is second-order accurate in the case of linear forcing. The values $\beta=1/4$ and $\eta=1/2$ lead to an unconditionally stable method  \cite{Newmark59}.

The Newmark method is written in the predicted-corrected form. The predicted values of $y_{n+1}$ and $\dot{y}_{n+1}$ are computed from the known values at time $t_n$:
\begin{equation}
\begin{array}{l}
\displaystyle
\tilde{y}_{n+1} = y_n + \Delta t\,\dot{y}_n + (1-2\beta)\,\frac{{\Delta t}^2}{2}\,\ddot{y}_n,\\
\\
\displaystyle
\tilde{\dot{y}}_{n+1} = \dot{y}_n + (1-\eta)\,\Delta t\,\ddot{y}_n.
\end{array}
\label{NewmarkP}
\end{equation}
The corrected values at time $t_{n+1}$ are
\begin{equation}
\begin{array}{l}
\displaystyle 
y_{n+1} = \tilde{y}_{n+1} + \beta\,{\Delta t}^2\,\ddot{y}_{n+1},\\
\\
\displaystyle
\dot{y}_{n+1} = \tilde{\dot{y}}_{n+1} + \eta\,\Delta t\,\ddot{y}_{n+1} .
\end{array}
\label{NewmarkC}
\end{equation}
To compute (\ref{NewmarkC}), one needs $\ddot{y}_{n+1}$. For this purpose, the corrected values (\ref{NewmarkC}) are injected into (\ref{OscMeca1}), yielding the displacement of the lip at time $t_{n+1}$:
\begin{equation}
\begin{array}{l}
\ds
\hspace{-0.5cm}
y_{n+1}=\tilde{y}_{n+1}\\
[8pt]
\ds +\beta\,\Delta t^2\,\frac{f\left(y_{n+1},t_{n+1}\right)-r\,\tilde{\dot{y}}_{n+1}-k\,(\tilde{y}_{n+1}-y_\textit{eq})}{m+r\,\eta\,\Delta t+k\,\beta\,{\Delta t}^2}.
\end{array}
\label{Ynp1}
\end{equation}
The aeroacoustic force $f(y,t)$ in (\ref{AeroForce})-(\ref{ExciterDepNL}) depends nonlinearly upon $y$. Consequently, the displacement $y_{n+1}$ in (\ref{Ynp1}) is the solution of the fixed point equation
\begin{equation}
g(z)=z,
\label{GzZ}
\end{equation}
with
\begin{equation}
\begin{array}{l}
\ds
\hspace{-0.5cm}
g(z)=\tilde{y}_{n+1}\\
[8pt]
\ds+\beta\,\Delta t^2\,\frac{f\left(z,t_{n+1}\right)-r\,\tilde{\dot{y}}_{n+1}-k\,(\tilde{y}_{n+1}-y_\textit{eq})}{m+r\,\eta\,\Delta t+k\,\beta\,{\Delta t}^2}.
\end{array}
\label{FixedPoint}
\end{equation}
A fixed-point method is used to solve (\ref{FixedPoint}). It is initialized by $y_n$, and then it is performed until the relative variation in (\ref{FixedPoint}) doesn't exceed $10^{-13}$. At each step of the fixed-point method, one takes $p_e^-(z,t_{n+1})=p_0^{(n+1)-}$: this value of the incoming pressure at node 0 and time $t_{n+1}$ is known, based on the propagation step in the resonator (section \ref{SecResoNumSchem}). The coefficient $\xi$ in (\ref{ExciterDepNL}) follows from (\ref{CoeffsCoupl1}).

Once the displacement $y_{n+1}$ is known, the acceleration is updated based on (\ref{NewmarkC}):
\begin{equation}
\ddot{y}_{n+1}=\frac{y_{n+1}-\tilde{y}_{n+1}}{\beta\,\Delta t^2}.
\label{UpdateNewmark}
\end{equation}
The velocity $\dot{y}_{n+1}$ is deduced from (\ref{NewmarkC}) and (\ref{UpdateNewmark}).


\subsubsection{Validation}\label{SecExcNumVal}

\begin{figure}[htbp]
\begin{center}
\begin{tabular}{c}
(i)\\
\includegraphics[scale=1.0]{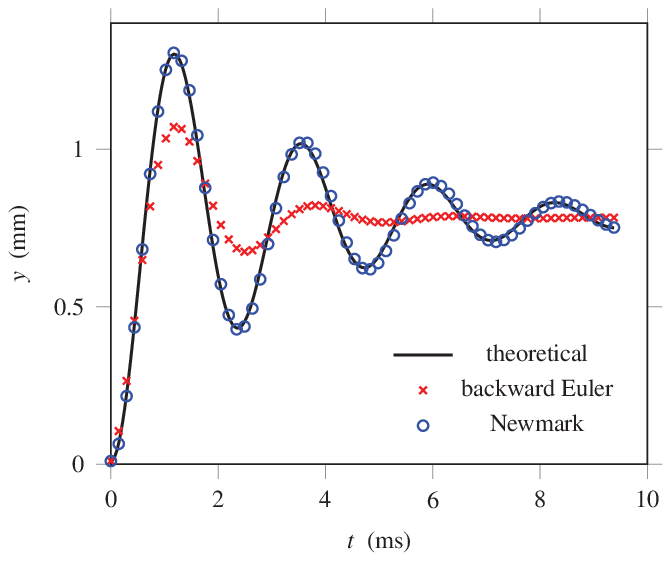} \\
(ii)\\
\includegraphics[scale=1.0]{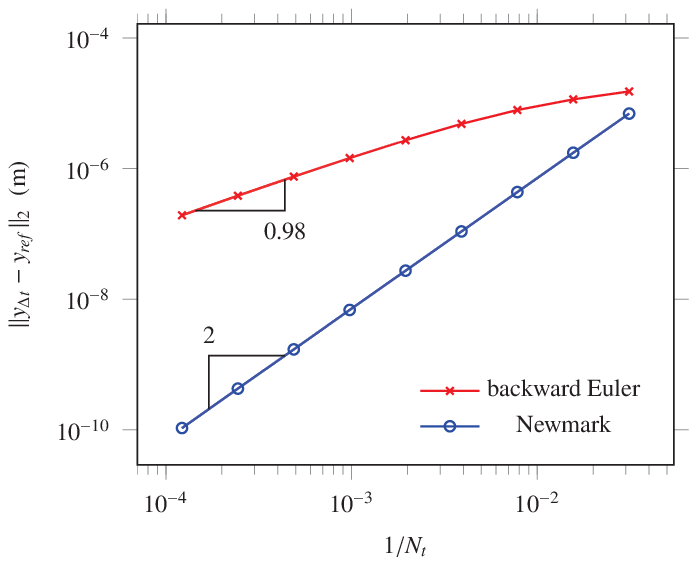} 
\end{tabular}
\caption{\label{FigOrdreSchema} Numerical resolution of the ordinary differential equation (\ref{OscMeca}) with a linear step forcing. (i): time histories of the numerical and exact solution; (ii): convergence measurements.}
\end{center}
\end{figure}

No closed-form solution of (\ref{OscMeca}) is known. To assess the accuracy of the Newmark method, we consider the  linear case of a step forcing: $f(y,t)=H(t)$. The parameters are those of table~\ref{TabParam2}, where the initial conditions are $y_0=y_\textit{eq}=0$~m and $y_1=0$~m/s. The numerical solution is computed on $N_t = 64$ time steps, up to 10~ms (here $\Delta t = 10/64$~ms is constant). Figure~\ref{FigOrdreSchema}-(i) compares the Newmark solution to the exact one. For completeness, the solution obtained by the backward Euler method is also displayed. Agreement is obtained between the Newmark solution and the exact one; on the contrary, the Euler solution suffers from a large numerical dissipation.

Measures of convergence are performed by considering various numbers of time steps, from $N_t = 32$ to $N_t = 8192$, and by computing the numerical solution up to 10~ms. The errors between the numerical solutions and the exact solution are displayed on figure~\ref{FigOrdreSchema}-(ii) in log-log scales. Second-order accuracy is obtained with Newmark's method, whereas only first-order acuracy is obtained with Euler's method.


\section{Numerical experiments}\label{SecExp}

\subsection{Summary of the algorithm}\label{SecExpAlgo}

Here we sum up the coupling between the resonator and the exciter. Time-marching from time $t_n$ to $t_{n+1}$ is as follows ($i=0,\cdots,N_x$):
\begin{enumerate}
\item Resonator
\begin{enumerate}
	\item computation of the outgoing and incoming velocities $u_{i>0}^{(n+1)+}$ and $u_{i<N_x}^{(n+1)-}$ using the numerical scheme (\ref{AlgoSplitting});
	\item computation of the incoming pressure at the input of the instrument $p_0^{(n+1)-}$ according to (\ref{SurP});
	\item update of $u_{N_x}^{(n+1)-}$ at $x=D$, according to the reflection condition (\ref{SystComplet4}).
\end{enumerate}
\item Exciter
\begin{enumerate}
	\item calculus of the lips opening $y_{n+1}$ in (\ref{NewmarkC}) based on the Newmark method and on the pressure at the entry of the resonator $p_e^{n+1}$ (\ref{ExciterDepNL});
	\item calculus of the outgoing pressure at the resonator's entry $p_0^{(n+1)+}=p_e^{n+1}-p_0^{(n+1)-}$  (\ref{Pe}), where $p_0^{(n+1)-}$ is known according to the step 1-(b) of the algorithm;
	\item update of the forcing source $u_0^{(n+1)+}$ in the resonator (\ref{SystComplet3}), based on $p_0^{(n+1)+}$ and (\ref{SurP}).
\end{enumerate}
\item Incrementation
\begin{enumerate}
	\item computation of the time step $\Delta t$, according to the CFL condition (\ref{CFL});
	\item affectation $n\gets n+1$.
\end{enumerate}
\end{enumerate}


\subsection{Configuration}\label{SecExpConf}

\begin{table}[htbp]
\caption{\label{TabParam2}Physical and geometrical parameters of the lips.}
\begin{center}
{\renewcommand{\arraystretch}{1.3}
\renewcommand{\tabcolsep}{0.3cm}
\begin{tabular}{c|c|c}
\hline
$m$ (kg)            & $k$ (N/m)   & $r$ (N.s/m)          \\
\hline  
$1.78\cdot 10^{-4}$ & $1278.8$    & $\sqrt{m\, k}/4$     \\
\hline\hline 
$l$ (m)             & $A$ (m$^2$) & $p_m$ (Pa)           \\
\hline
$10^{-2}$           & $10^{-4}$   & $20\cdot 10^{3}$     \\
\hline\hline
$y_0$ (m)           & $y_1$ (m/s) & $y_{\it eq}$ (m)\\
\hline
$4\cdot 10^{-3}$    & $-4$        & $5\cdot 10^{-4}$     \\
\hline
\end{tabular}}
\end{center}
\end{table}

The wave propagation is described by the complete Menguy-Gilbert model in a resonator with a constant radius $R=7$~mm. The distance $D=1.4$~m from the input to the output of the resonator is discretized on $N_x=100$ points. 
The parameters of the lips model are given in table \ref{TabParam2}. These parameters are issued both from different publications  \cite{adachi2, vilain2003a} and from trial and errors until self-oscillations are obtained.

The output of the model is the acoustic velocity at the end of the tube $u(D,t)=u^+(D,t)+u^-(D,t)$. Considering that the open end of the cylinder radiates as a monopole, $u(D,t)$ is converted into $p_{rec}(t)$, the pressure measured at an arbitrary distance $D_{rec}=10$~m from the output of the cylinder, through the relation:
\begin{equation}
p_{rec}(t) = \frac{\rho_0\,S}{4\,\pi\,D_{rec}} \frac{\partial u}{\partial t}(D,t).
\end{equation}
This model could be obviously refined : since radiation only acts as a boundary condition, it is completely independant of the propagation inside the waveguide which is the focus of this paper.


\subsection{Results}\label{SecExpRes}

Various numerical experiments are carried out in order to check the influence of the nonlinear wave propagation on the behavior of the model. Simulations are carried out on Scilab and last 18 minutes for each computed second, when a mid-range laptop is used (Intel Core i7, 2.4 GHz, 8 Go, 2011).  Time domain signals $p_{rec}(t)$ presented in the following figures are normalized by their maximal value. Originally registered with a variable time step, they are then sampled at a frequency $f_{s}=44.1$~kHz, by the use of linear interpolation.

\begin{figure}
\includegraphics[width=1.05\columnwidth]{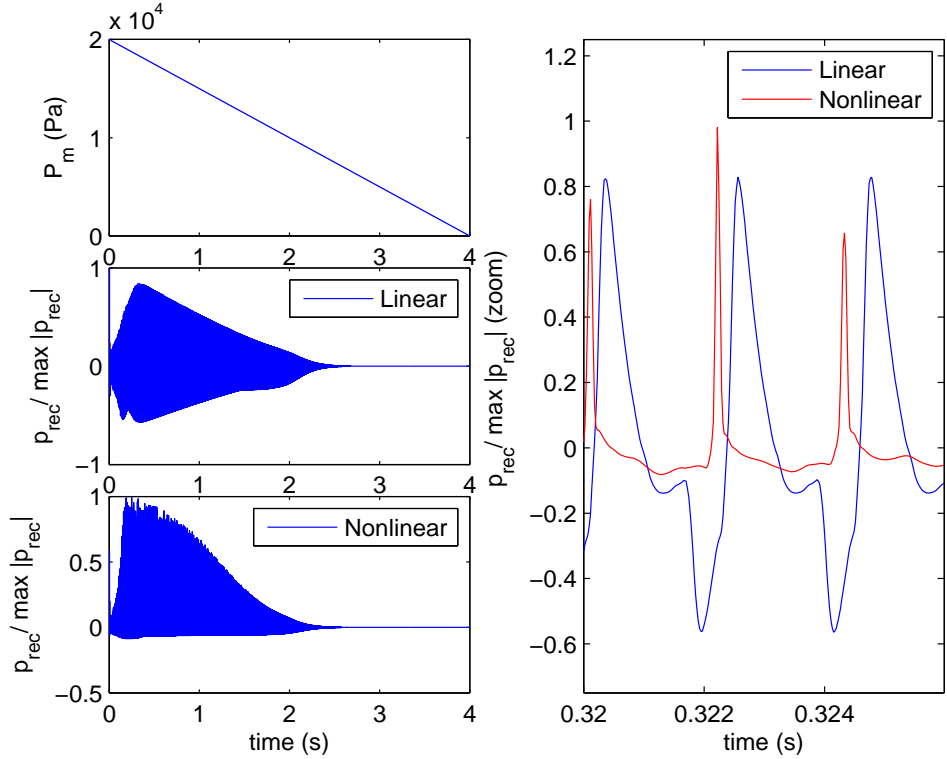}
\caption{Progressive decrease of the blowing pressure, from $p_m=20$~kPa downto $p_m=0$~kPa. Left: time histories of $p_m$ (top), $p_{rec}/\!\max |p_{rec}|$ with linear wave propagation (middle), and with nonlinear wave propagation (bottom). Right: zoom on a few periods of $p_{rec}/\!\max |p_{rec}|$ in the linear and nonlinear cases.
\label{f:RampDesc_sig}}
\end{figure}

In figure~\ref{f:RampDesc_sig}, the blowing pressure $p_m(t)$ decreases during 4 s, from $p_m=20$~kPa downto $p_m=0$~kPa. Two regimes are considered: linear wave propagation ($b=0$), and nonlinear wave propagation. The recorded pressure $p_{rec}$ displays radically different time envelopes in the two cases (left column, middle and bottom): a shortest attack transient in the linear case, and an extinction threshold occurring for lower $p_m$ in the nonlinear case. In the linear case, the signal is more symmetric with respect to zero. Moreover, the regime is slightly quasi-periodic in the nonlinear case for high oscillating amplitudes. A closer view on the signals (right column) reveals typical waveforms with sharp peaks in the nonlinear case. 

\begin{figure}
\includegraphics[width=1.05\columnwidth]{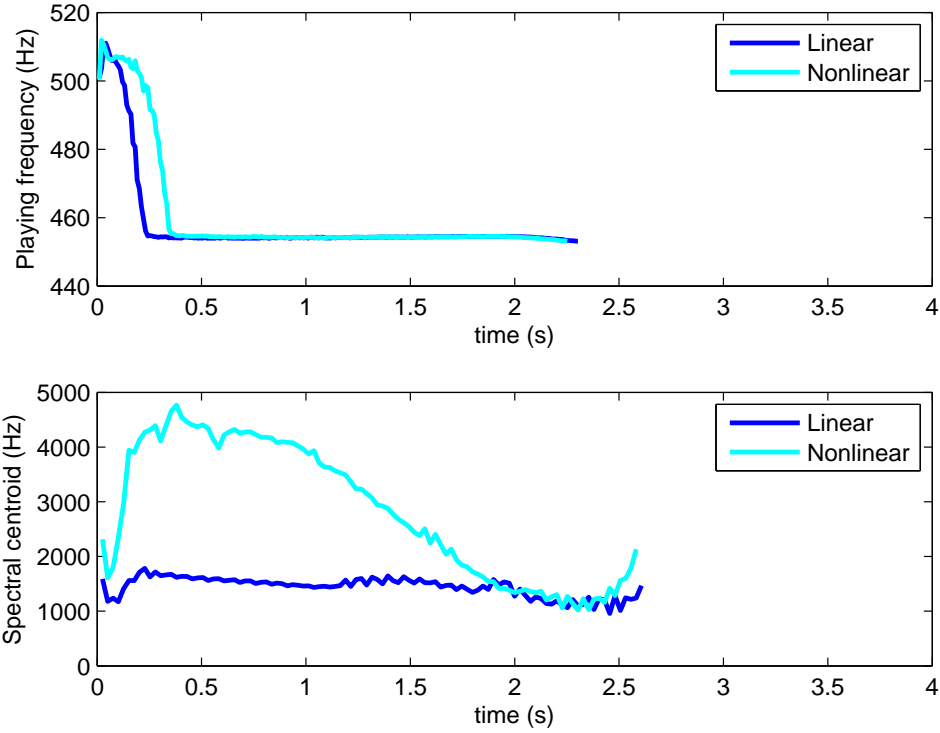}
\caption{Descriptors calculated with the MIR Toolbox  \cite{mirtoolbox} according to the time domain signals presented in figure~\ref{f:RampDesc_sig}. Top: playing frequency; bottom: spectral centroid.
\label{f:RampDesc_ind}}
\end{figure}

Based on these data, two descriptors are computed in the frequency domain and displayed in figure~\ref{f:RampDesc_ind}: the playing frequency (top) and the spectral centroid (bottom). During an initial phase, playing frequencies differ significantly  between the linear and nonlinear cases: the instrument plays at higher frequencies if nonlinear propagation is taken into account, up to $45$~Hz at most around $t=0.25$~s (+157 cents, i.e. between half a tone and a tone). The influence of nonlinear propagation on the playing frequency vanishes around $t=0.35$~s. It is worth noting that even at high oscillating amplitude, a negligible difference is observed after $t=0.35$~s: only 3 cents around $t=0.5$~s. It is also striking that attack time (defined here as the time for the signal to reach the maximum amplitude from $t=0$~s) is almost twice as long in the linear case than in the nonlinear case ($0.35$~s versus $0.18$~s, see figure \ref{f:RampDesc_sig}). However the time during which the playing frequency varies significantly is much longer in the nonlinear case ($0.35$~s versus $0.24$~s, see figure \ref{f:RampDesc_ind}). 

The influence of the nonlinear propagation on the playing frequency had already been highlighted numerically in the case of the trombone,  \cite{msallam00} but only for steady states regimes, yet with lower blowing pressures and a simplified model for the nonlinear propagation. Deviations of less than 5 cents for weak dynamics have been reported and are in agreement with our observations.  But the picture is very different during the transient phase, as explained above. 

 The bottom picture of figure~\ref{f:RampDesc_ind} confirms that the nonlinear propagation is associated with an enrichment of the sound spectrum with high frequencies. Indeed the spectral centroid is up to three times higher than in the case of linear propagation. Moreover in the case of linear propagation, the spectral centroid is nearly constant, which suggests that the nonlinearity due to the exciter  (\ref{ExciterDepNL}) cannot explain the spectral enrichment features of brassy sounds. On the contrary, in the case of nonlinear propagation, the spectral centroid is a monotonic function of the oscillating amplitude.

\begin{figure}
\includegraphics[width=1.05\columnwidth]{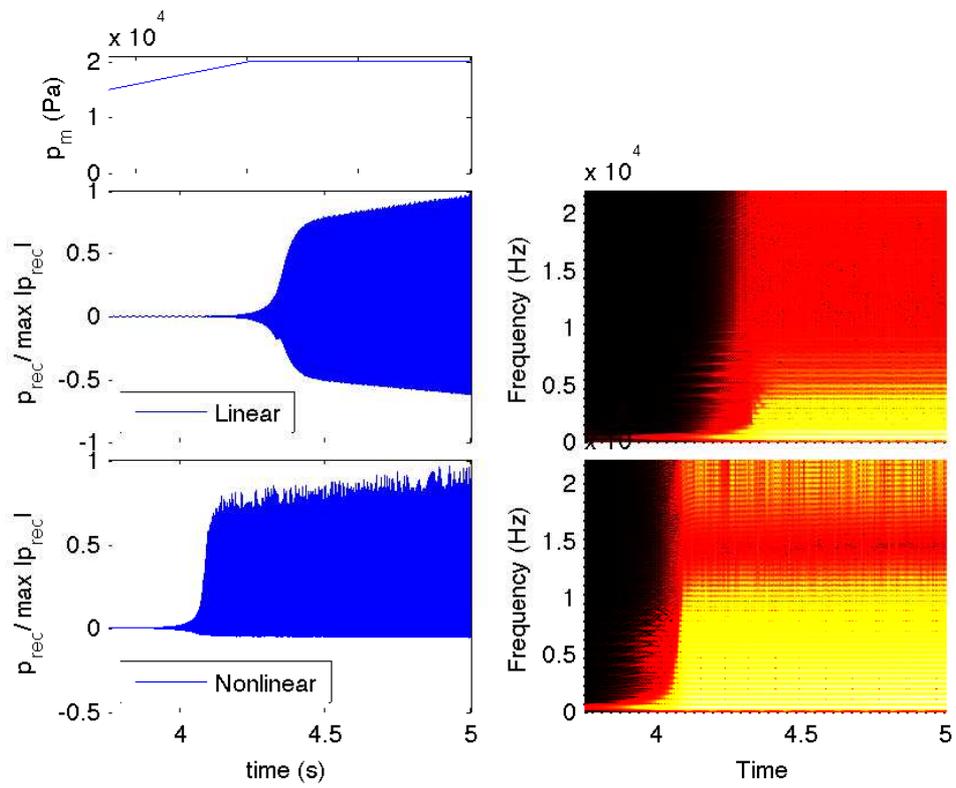}
\caption{Progressive increase of the blowing pressure from $p_m=0$~kPa to $p_m=20$~kPa during $5$~s 
(only a zoom is shown between $t=3.75$~s and $t=5$~s). Left: time histories of $p_m$ (top), $p_{rec}/\!\max |p_{rec}|$ with linear wave propagation (middle) and  with nonlinear wave propagation (bottom). Right: spectrogram of $p_{rec}/\!\max |p_{rec}|$ in the case of linear (middle) and nonlinear wave propagation (bottom).
\label{f:RampMont_sig.eps}}
\end{figure}

Another numerical experiment is carried out by linearly increasing $p_m$ from $p_m=0$~kPa to $p_m=20$~kPa during 5~s, 
Time domain signals are presented in figure~\ref{f:RampMont_sig.eps} (left column). The most striking feature is the fact that  the oscillation threshold is shifted toward higher values of $p_m$ when the nonlinear propagation is ignored ($b=0$ in (\ref{SystComplet1})). One should speak preferably about ``dynamic oscillation threshold"  \cite{bergeot2013a} instead of ``oscillation threshold" since these observations are done while the bifurcation parameter (here the blowing pressure $p_m$) is being varied in time. This result is surprising at first glance since the consequences of the nonlinear propagation are expected to vanish at the oscillation threshold where $b\frac{(u^\pm)^2}{2} \ll a|u^\pm|$ in (\ref{SystComplet1}). However, the behavior of dynamic bifurcations thresholds can be counterintuitive, even when considering small perturbations  \cite{bergeot2013b}. The so-called bifurcation delay observed here is around $0.2$~s, which corresponds to a pressure difference around $800$~Pa. In the right column of figure~\ref{f:RampMont_sig.eps}, the spectrograms of the time signals highlight two major features: first, the signal calculated while considering nonlinear propagation has a much more broadband structure. Secondly, the spectral content with the amplitude of the signal evolves more significantly than in the hypothesis of linear propagation. This is consistent with experimental observations in brass instruments  \cite{hirschberg96b, rendon2013}. 

\begin{figure}
\includegraphics[width=1.05\columnwidth]{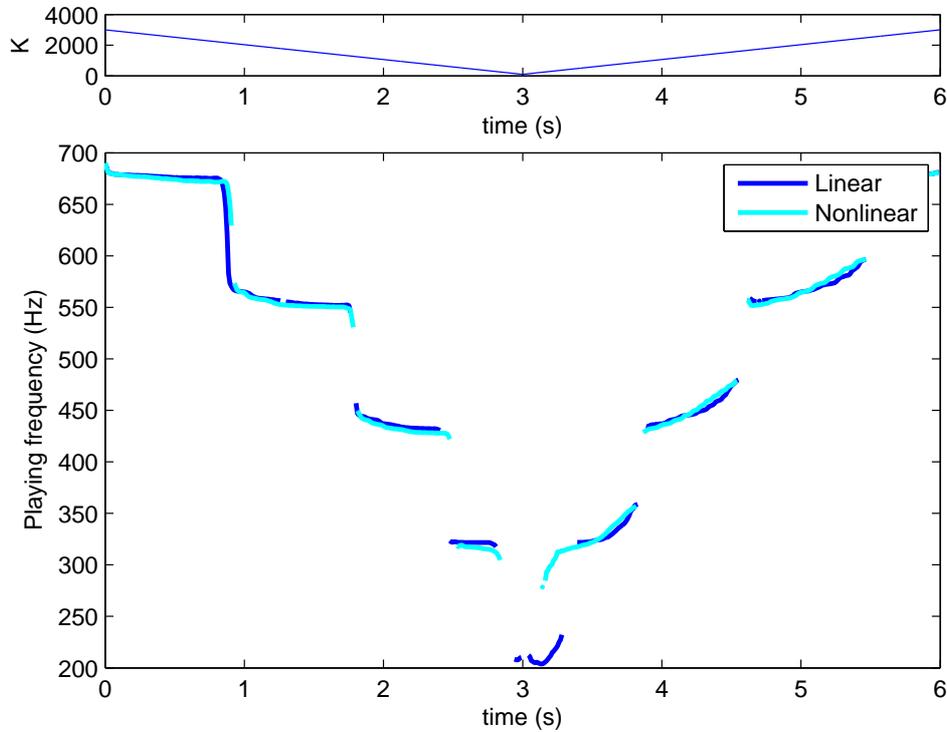}
\caption{Symmetric decrease / increase of the stiffness of the lips model between $k=3000$~N.m$^{-1}$ and $k=100$~N.m$^{-1}$. The blowing pressure is constant: $p_m=20$~kPa. Top: time history of $k$. Bottom: time history of the playing frequency with linear wave propagation (blue) and under nonlinear wave propagation hypothesis (red).
\label{f:FreqVarTemps}}
\end{figure}

In a third experiment, a constant blowing pressure $p_m=20$~kPa is considered. The stiffness $k$ of the lip model follows a symmetric decrease / increase during 6~s between $k=3000$~N.m$^{-1}$ and $k=100$~N.m$^{-1}$, as shown in figure~\ref{f:FreqVarTemps} (top). As expected, the model plays on different periodic regimes (corresponding to the $2^{nd}$ to the $6^{th}$ registers), the frequencies of which are displayed in the bottom picture. The most striking result is that for the parameters values chosen for the simulation, the lowest register is not playable in the case of nonlinear propagation. A closer view reveals that for each register, frequency jumps with the neighboor registers (lower and upper) do not occur at the same thresholds. Concerning the playing frequencies, differences may be weak but clearly audible and differences are all the larger as the playing frequency (i.e. the register) is low:  up to $10$~cents on the $6^{th}$  register, up to $11.5$~cents on the $5^{th}$, up to $16$~cents on the $4^{th}$, up to $36$~cents on the $3^{rd}$. The playing frequency is always lower in the case of nonlinear propagation when $k$ is decreased. When $k$ is increased, for each register, the playing frequency is lower in the case of nonlinear propagation during the first half of the register. However, since it increases faster than in the case of linear propagation,  the playing frequency in the case of nonlinear propagation becomes higher in the second half of the register. Here again, considering nonlinear propagation appears to have a noticeable effect during transients of a control parameter.

\begin{figure}
\includegraphics[width=1.05\columnwidth]{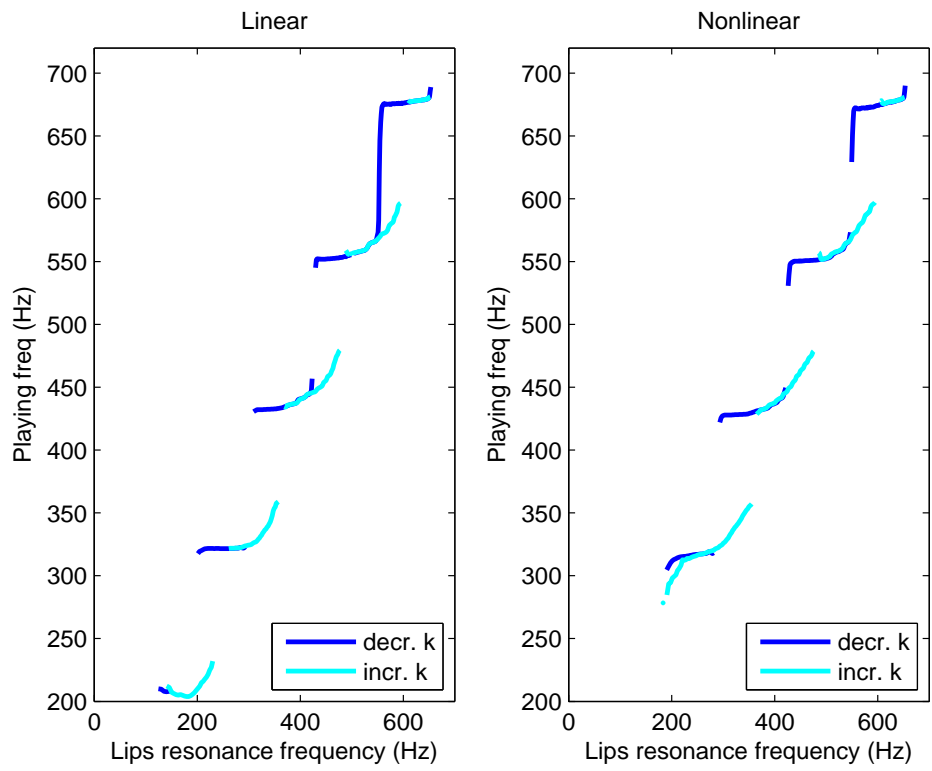}
\caption{Same data than in figure~\ref{f:FreqVarTemps}. Playing frequency is plotted with respect to the lips resonance frequency $\frac{1}{2\pi}\sqrt{\frac{k}{m}}$  with linear propagation ($b=0$ in (\ref{SystComplet1})) (left) and with nonlinear propagation hypothesis (right). 
\label{f:FreqVarFreq}}
\end{figure}

In order to highlight hysteresis effects, the same data is plotted with respect to the resonance frequency of the lips model in figure~\ref{f:FreqVarFreq} for linear (left) and nonlinear propagation (right). Hysteresis in such experiments is known to result from two mechanisms: the coexistence of stable periodic regimes and the variation with time of the bifurcation parameter (dynamic bifurcations).  The left part of the figure shows familiar simulation results  \cite{silva2014a}. Considering the nonlinear propagation does not alter significantly  the hysteresis excepted for the $3^{rd}$ register (the lowest on the right picture). In this case, a larger hysteresis is observed in the case of nonlinear propagation, which is possibily linked to the fact that the model failed to produce the $2^{nd}$ register. 


\section{Conclusion}\label{SecConclu}

A time-domain numerical modeling of brass instruments has been proposed. The propagation of outgoing and incoming nonlinear acoustic waves has been considered, taking into account the viscothermal losses at the boundaries of the resonator. The coupling with a model of lips has been modeled also, enabling to simulate the self-sustained oscillations in brass instruments. The software so-obtained has been extensively validated. Preliminary applications to configurations of interest in musical acoustics have been demonstrated.

In its current form, our simulation tool can be used to investigate various open questions in acoustics. The first one concerns the frequency response of a nonlinear acoustical resonator, which has already been the subject of experimental and theoretical works \cite{Ilinskii98,Hamilton01}. For this purpose, the methodology followed to determine the linear impedance (section \ref{SecResoValImp}) can be adapted to the nonlinear regime. A second application is to study numerically the threshold of oscillations in brass instruments. Based on a modal representation of the field in the resonator, a Floquet theory can be applied in the linear regime  \cite{Ricaud09}. But to our knowledge, no results are known when the nonlinearity of the wave propagation is taken into account. On the contrary, the numerical tool does not suffer from such a limitation.

The physical modeling has also to be improved. In particular, considering simple outgoing and incoming waves is a crude assumption in a tube with a variable section. In the linear regime of propagation, the Webster-Lokshin wave equation provides a more realistic framework \cite{Haddar10}. Extension of this equation to the nonlinear regime of propagation has been considered by Bilbao and Chick, \cite{Bilbao13} but without the viscothermal losses. The derivation of the full bidirectional system---incorporating nonlinear wave propagation, and viscothermal losses in a variable tube---is a subject of current research.


\end{document}